\documentclass[letterpaper, preprint, paper, 11pt]{AAS}	

\usepackage[utf8]{inputenc}
\usepackage{bm}
\usepackage{amsmath}
\usepackage{amssymb}
\usepackage[colorlinks=true, pdfstartview=FitV, linkcolor=black, citecolor=black, urlcolor=black]{hyperref}
\usepackage{overcite}
\usepackage{footnpag}			      	
\usepackage[capitalise,noabbrev]{cleveref}

\newcommand{\mb}[1]{\boldsymbol{#1}}

\PaperNumber{26-844}

\begin{document}

\title{LQR Design for Formation Flying Near Halo Orbits Exploiting Quasi-Periodic Symmetry in Toroidal Coordinates}

\author{Joaquin G. Lopez-Cepero\thanks{Ph.D. Candidate, Department of Aerospace Engineering and Fluid Mechanics, University of Seville, Seville, Spain, jgonzalez47@us.es.},
Rafael Vazquez\thanks{Full Professor, Department of Aerospace Engineering and Fluid Mechanics, University of Seville, Seville, Spain, rvazquez1@us.es.},
\ and Julio C. Sanchez\thanks{Associate Professor, Department of Aerospace Engineering and Fluid Mechanics, University of Seville, Seville, Spain, jsanchezm@us.es.}}

\maketitle

\begin{abstract}
This work presents the design of a Linear Quadratic Regulator for relative motion control near periodic orbits in the Circular Restricted Three-Body Problem (CR3BP), formulated in non-singular toroidal coordinates. The key result exploits the rotational quasi-periodicity of the toroidal coordinate transformation. A uniqueness argument on the stabilizing solution of the associated difference Riccati equation proves the optimal control gains need only be computed for a single orbital period. The controller is validated against the full non-linear CR3BP dynamics on an L1 Northern halo orbit in the Earth-Moon system, and further against a high-fidelity ephemeris model, demonstrating successful reconfiguration maneuvers on the invariant torus with low control effort. Compared with an impulsive targeting and station-keeping baseline, the proposed controller attains a comparable tracking accuracy while reducing the total control effort by approximately 40\%.
\end{abstract}

\section{Introduction}

The cislunar environment has recently emerged as a highly strategic domain, both as a focus of continued lunar exploration and as a proving ground for broader Solar System missions \cite{whitley2016options}. The planned deployment of permanent infrastructure, such as the Lunar Gateway, necessitates the development of robust astrodynamics frameworks to support autonomous operations, rendezvous, and formation flying in a highly non-linear multi-body gravitational regime. Within this context, Near Rectilinear Halo Orbits (NRHOs) (a specific subset of the L1 and L2 halo families) have been identified as primary candidates for long-term operational staging \cite{howell2018nrho}.

In the Circular Restricted Three-Body Problem (CR3BP) \cite{szebehely1967theory}, periodic orbits can exhibit an oscillatory mode, indicated by a pair of complex conjugate eigenvalues lying on the unit circle. In the vicinity of such orbits, the natural relative dynamics are governed by nearby invariant 2-tori, upon which a spacecraft will trace out bounded, quasi-periodic trajectories \cite{capannolo2023model}. These natural quasi-periodic relative orbits are ideal for formation flying, as they allow multiple spacecraft to maintain bounded relative configurations with very low control effort \cite{elliott2022describing}.

Designing controllers that exploit this toroidal structure poses two main challenges. First, numerically computing the full non-linear invariant tori requires expensive continuation or collocation methods and produces high-dimensional objects that are impractical to store or evaluate onboard \cite{capannolo2023model}. To address this, Reference~\citenum{elliott2022describing} proposes using a first-order approximation of the invariant 2-torus, constructed from the center-mode eigenvector of the monodromy matrix, to define a local, non-singular toroidal coordinate frame. This approximation requires only the State Transition Matrix over one orbital period. Second, even when using the proposed non-singular toroidal coordinates, the resulting relative dynamics are non-periodic Linear Time-Varying (LTV), which precludes the standard constant-gain infinite-horizon Linear Quadratic Regulator (LQR) solution obtained from a single algebraic Riccati equation.

Control techniques in the vicinity of periodic orbits have been explored in several works. In Reference~\citenum{peng2011optimal}, a periodic LQR is designed for formation flying and station-keeping by linearizing the dynamics around the periodic orbit. In Reference~\citenum{lopez2022control}, a periodic LQR is obtained by exploiting the Lyapunov-Floquet transformation. A robust Model Predictive Control (MPC) for rendezvous operations has also been developed in Reference~\citenum{sanchez2020chance}.

Moreover, control in the vicinity of invariant tori has been addressed through MPC in the rotating frame \cite{capannolo2023model} and impulsive controllers formulated for the non-singular toroidal coordinates \cite{elliott2021impulsive}. The use of toroidal coordinates allows for an intuitive design of formation flying strategies exploiting quasi-periodic orbits.

This work proposes the design of an LQR controller exploiting specific properties of the non-singular toroidal coordinate transformation that enables the formulation of an infinite-horizon controller fully designed within the non-singular toroidal frame. In particular, a state weight matrix that is invariant with respect to the quasi-periodic symmetry of the toroidal coordinates is used. With this methodology, it is possible to obtain optimal gains that can be stored compactly and extended online via a trivial rotation. 

\section{Dynamical model}

The design of the reference trajectories is done using the CR3BP \cite{capannolo2023model}, which describes the motion of a massless particle under the gravitational influence of two primary bodies in circular orbits around their common center of mass. A rotating frame of reference is used, where the two primary bodies are fixed on the $x$-axis \cite{lopez2022control}, which rotates with an angular velocity $\omega_z$ around its $z$-axis. The equations of motion in the rotating frame are given by

\begin{equation}
    \ddot{x} - 2\dot{y} = \dfrac{\partial \Omega}{\partial x}, \quad \ddot{y} + 2\dot{x} = \dfrac{\partial \Omega}{\partial y}, \quad \ddot{z} = \dfrac{\partial \Omega}{\partial z},
\end{equation}

where $\Omega$ is the pseudo-potential function defined as
\begin{equation}
    \Omega(x,y,z) = \dfrac{1}{2}(x^2+y^2) + \dfrac{1-\mu}{r_1} + \dfrac{\mu}{r_2},
\end{equation}

 and $\mu = m_2/(m_1+m_2)$ is the mass ratio of the system. $r_1$ and $r_2$ are the distances from the particle to the two primary bodies, given by
\begin{equation}
    r_1 = \sqrt{(x+\mu)^2 + y^2 + z^2}, \quad r_2 = \sqrt{(x-1+\mu)^2 + y^2 + z^2}.
\end{equation}

The equations of motion are non-dimensionalized by using the following characteristic units: 
\begin{equation}
    L^* = |\mb{r}_2 - \mb{r}_1|, \quad T^* = 1/\omega_z, \quad M^* = m_1 + m_2.  
\end{equation}

The mass ratio for the Earth-Moon system is $\mu_{EM} \simeq 1.2150585\times 10^{-2}$.

\section{Non-singular toroidal coordinates definition}
To establish the relative motion framework used in this study, we adopt the non-singular toroidal coordinates formalized in Reference~\citenum{elliott2022describing}.

Let $\mb{\Phi}(t;t_0)$ be the State Transition Matrix (STM) from time $t_0$ to time $t$ associated with a reference periodic orbit of period $T$ within the CR3BP. Mathematically, the monodromy matrix $\mb{M}(t) := \mb{\Phi}(t+T;t)$ evaluated at a state $\mb{x}(t)$ along the periodic orbit possesses at least one pair of trivial eigenvalues equal to one. If the orbit admits an oscillatory mode, the remaining spectrum includes a complex conjugate pair on the unit circle, expressed as $\lambda = e^{\pm i \omega}$ with $\omega \in \mathbb{R}$. As shown in Reference~\citenum{elliott2022describing}, the complex eigenvector $\mb{w}$ associated with this oscillatory mode provides the basis vectors that span the local center eigenspace.

As shown in Reference~\citenum{elliott2022describing}, a first-order approximation of an invariant curve tangent to the invariant torus can be expressed as a function of the complex eigenvector $\mb{w}$ associated with the oscillatory mode as: 
\begin{equation}
    \mb{\varphi} = \epsilon \left(\text{Re}(\mb{w})\cos \theta + \text{Im}(\mb{w})\sin\theta\right) \quad \theta \in [0,2\pi],
\end{equation}

where $\epsilon$ is a scaling parameter. For a fixed $\epsilon$, the real and imaginary parts of the eigenvector, $\mb{w}_r$ and $\mb{w}_i$, define two conjugate diameters of the ellipse traced by the states in $\mb{\varphi}$, which is tangent to the invariant torus.

To fully construct the torus, $\mb{w}$ must be computed at each time instant during one full revolution, which can be done by using the STM to map the eigenvector to each time instant.

Due to the periodicity of the orbit, the STM and the eigenvector $\mb{w}$ need only be computed for a single period. The values for $t>T$ can be obtained by exploiting the property of the STM on periodic orbits, where $\mb{M} = \mb{\Phi}(t_0+T;t_0)$ is the monodromy matrix:
\begin{equation}
    \mb{\Phi}(t+nT;t_0) = \mb{\Phi}(t;t_0)\mb{M}^n.
    \label{eq:STMidentity}
\end{equation}

The complex eigenvector referring to the center mode may be written as
\begin{equation}
    \mb{w} = \begin{bmatrix}
        \mb{r}_r \\ \mb{v}_r
    \end{bmatrix} + \begin{bmatrix}
        \mb{r}_i \\ \mb{v}_i
    \end{bmatrix}i.
\end{equation}

At each time instant, $\{\mb{r}_r, \mb{r}_i\}$ define the plane that contains the projection of the center eigenspace onto the position domain. The normal vector to this plane is
\begin{equation}
    \hat{\mb{n}} = \dfrac{\mb{n}}{n} = \dfrac{\mb{r}_r \times \mb{r}_i}{|\mb{r}_r \times \mb{r}_i|}.
\end{equation}

This eigenvector may be normalized in such a way that the directions defined by $\mb{r}_r$ and $\mb{r}_i$ coincide with the semi-major and semi-minor axes of the projection of the invariant curve onto the position domain (see, e.g. \cite{elliott2022describing}). This normalization is applied only at the reference epoch $t_0$; the frame at any other time is obtained by propagating $\mb{w}$ with the STM, so that over one period the basis $\{\mb{r}_r, \mb{r}_i\}$ returns rotated by $\omega$ rather than repeating. This rotation is precisely what is captured by $\mb{\Gamma}$ in the section on the time evolution of the transformation matrix.

The vectors $\{\mb{r}_r, \mb{r}_i, \hat{\mb{n}}\}$ define the non-singular toroidal frame $\mathcal{Z}$ at each time instant.

Consider a chaser spacecraft in the vicinity of the reference periodic orbit. Its relative position in the non-singular toroidal coordinates is given by $\mb{z} = [\alpha, \beta, h]^T$. The first two coordinates are the projections of the relative position onto the plane defined by $\mb{r}_r$ and $\mb{r}_i$. The last coordinate is the projection of the relative position onto the normal vector $\hat{\mb{n}}$.

$\mb{z}$ is related to the rotating-frame relative position by $\delta\mb{r} = \mb{R}\mb{z}$, where $\mb{R} = [\mb{r}_r, \mb{r}_i, \hat{\mb{n}}]$ is a time-varying, generally non-orthogonal matrix. As long as $\mb{r}_r$ and $\mb{r}_i$ are not aligned (i.e. the oscillatory mode does not produce rectilinear motion), $\mb{R}$ is invertible and the toroidal coordinates are obtained as $\mb{z} = \mb{R}^{-1}\delta\mb{r}$ \cite{elliott2022describing}. 

The time derivatives of the toroidal coordinates and the full transformation between the toroidal state $\mb{Z} = [\mb{z}; \dot{\mb{z}}] = [\alpha, \beta, h, \dot{\alpha}, \dot{\beta}, \dot{h}]^T$ and the rotating-frame relative state $\delta \mb{x} = [\delta \mb{r}; \delta \mb{v}]$ follow from differentiating this relation (see \cite{elliott2022describing} for details). The resulting transformation matrix is

\begin{equation}
    \mb{T} = \begin{bmatrix}
        \mb{R} & \mb{0} \\
        \dot{\mb{R}} & \mb{R}
    \end{bmatrix},
\end{equation}

where $\dot{\mb{R}} = [\mb{v}_r, \mb{v}_i, \dot{\hat{\mb{n}}}]$, with $\dot{\hat{\mb{n}}}$ computed from the cross-product differentiation of $\mb{r}_r \times \mb{r}_i$ (see \cite{elliott2022describing}).

By exploiting the matrix $\mb{T}$, the STM in the toroidal frame ${}^{\mathcal{Z}}\mb{\Phi}(t;t_0)$ may be computed by transforming $\mb{\Phi}(t;t_0)$ as follows:
\begin{equation}
    {}^{\mathcal{Z}}\mb{\Phi}(t;t_0) = \mb{T}^{-1}(t)\mb{\Phi}(t;t_0)\mb{T}(t_0).
    \label{eq:STMtoroidal}
\end{equation}

\section{Discrete-time linear dynamics on the non-singular toroidal frame}

Let $k$ denote the discrete time instant associated with epoch $t_k$, where $t_k = t_0 + k\Delta t$. Let $t_{N_p} = T$ be the end time of the period. The STM from instant $t_k$ to $t_{k+1}$ in the toroidal frame is given by
\begin{equation}
\begin{split}
     {}^{\mathcal{Z}}\mb{\Phi}_k^{k+1} &={}^{\mathcal{Z}}\mb{\Phi}(t_{k+1};t_k) = {}^{\mathcal{Z}}\mb{\Phi}(t_{k+1};t_0)\left({}^{\mathcal{Z}}\mb{\Phi}(t_k;t_0)\right)^{-1} =\\
     &= \mb{T}_{k+1}^{-1}\mb{\Phi}_0^{k+1}\mb{T}_{0}\left(\mb{T}_{k}^{-1}\mb{\Phi}_0^{k}\mb{T}_{0}\right)^{-1} = \mb{T}_{k+1}^{-1}\mb{\Phi}_0^{k+1}\left(\mb{\Phi}_0^{k}\right)^{-1}\mb{T}_k.
\end{split}
\end{equation}

Note that, once the STM is computed for a full period offline, it is possible to compute $\Phi_{k}^{k+1}$ for $k>N_p$ by exploiting the property of the STM on periodic orbits given by \cref{eq:STMidentity}.

Let $\mb{\xi} = [\delta \alpha, \delta \beta, \delta h, \delta \dot{\alpha}, \delta \dot{\beta}, \delta \dot{h}]^T$ be the relative state vector in the toroidal frame with respect to a (constant) reference $\mb{Z}_{ref}$. The relative linear discrete-time impulsive dynamics between instants $k$ and $k+1$ in the toroidal frame may be written as:
\begin{equation}
    \mb{\xi}_{k+1} = {}^{\mathcal{Z}}\mb{\Phi}_k^{k+1}\mb{\xi}_k+ \mb{B}_k\mb{u}_k = \mb{A}_k\mb{\xi}_k + \mb{B}_k\mb{u}_k,
    \label{eq:dynamics}
\end{equation}
where the input matrix $\mb{B}_k$ is defined as

\begin{equation}
    \mb{B}_k = \mb{A}_k\begin{bmatrix}
        \mb{0} \\
        \mb{R}^{-1}_k
    \end{bmatrix} = \mb{A}_k\mb{T}^{-1}_k\begin{bmatrix}
        \mb{0} \\
        \mb{I}
    \end{bmatrix}.
\end{equation}

Here, the control $\mb{u}_k$ is the impulsive velocity increment ($\Delta\mb{v}$) applied at instant $t_k$ and expressed in the rotating frame, which $\mb{B}_k$ maps into the toroidal relative state. These dynamics describe the evolution of the relative state in the toroidal frame under the effect of discrete-time control inputs. Note that, while the STM in the rotating frame is periodic, the STM in the toroidal frame is not, due to the non-periodicity of the transformation matrix $\mb{T}$. As a consequence, the presented system is LTV and not periodic.

\section{Time evolution of the transformation matrix}
\label{sec:TperiodicitySection}
Even though the transformation defined by $\mb{T}$ is non-periodic, let us analyze more closely the evolution of the columns of $\mb{R}$, which are given by the vectors $\mb{r}_r$, $\mb{r}_i$ and $\hat{\mb{n}}$.

Given that $\mb{w}$ is an eigenvector of the monodromy matrix, it satisfies the following relation:
\begin{equation}
    \mb{\Phi}(t+T;t)\mb{w}(t) = |\lambda|e^{i\omega}\mb{w}(t).
\end{equation}

Since the eigenvalues associated with the oscillatory mode form a complex conjugate pair of unit modulus, the previous equation can be rewritten as
\begin{equation}    \mb{\Phi}(t+T;t)\mb{w}(t) = e^{i\omega}\mb{w}(t).
    \label{eq:eigenvaluerelation}
\end{equation}

It can be proven that the normal vector $\hat{\mb{n}}$ is periodic, since it is given by the cross product of $\mb{r}_r$ and $\mb{r}_i$:

\begin{equation}
    \hat{\mb{n}}(t+T) = \hat{\mb{n}}(t).
\end{equation}

On the other hand,  \cref{eq:eigenvaluerelation} indicates that the real and imaginary components of $\mb{w}(t+T)$ are given by a rotation of the corresponding components of $\mb{w}(t)$.  As a consequence, the vectors $\mb{r}_r$ and $\mb{r}_i$ are not periodic, but it is possible to compute their values at time $t+T$ by applying a rotation. Let $\mb{R}_{\theta_\lambda}$ be the following rotation matrix of angle $\theta_\lambda = \omega$

\begin{equation}
    \mb{R}_{\theta_\lambda} = \begin{bmatrix}
        \cos\theta_\lambda & \sin\theta_\lambda & 0 \\
        -\sin\theta_\lambda & \cos\theta_\lambda & 0 \\
        0 & 0 & 1
    \end{bmatrix}.
\end{equation}

Then, the following relation holds:
\begin{equation}
    \mb{R}(t+T) = \mb{R}(t)\mb{R}_{\theta_\lambda}.
    \label{eq:Rperiodicity}
\end{equation}

From \cref{eq:Rperiodicity}, it is possible to deduce that
\begin{equation}
    \mb{T}(t+T) = \begin{bmatrix}
        \mb{R}(t)\mb{R}_{\theta_\lambda} & \mb{0} \\
        \dot{\mb{R}}(t)\mb{R}_{\theta_\lambda} & \mb{R}(t)\mb{R}_{\theta_\lambda}
    \end{bmatrix} = \mb{T}(t)\begin{bmatrix}
        \mb{R}_{\theta_\lambda} & \mb{0} \\
        \mb{0} & \mb{R}_{\theta_\lambda}
    \end{bmatrix} = \mb{T}(t)\mb{\Gamma}.
    \label{eq:TransformationPeriodicity}
\end{equation}

Note that $\mb{\Gamma}$ is an orthogonal matrix, $\mb{\Gamma}^T\mb{\Gamma} = \mb{I}$.

\subsection{Evaluation of the STM in toroidal frame for t greater than T}

By using \cref{eq:TransformationPeriodicity}, it is possible to compute $\mb{\Phi}_{k+N_p}^{k+N_p+1}$  as a function of $\mb{\Phi}_{k}^{k+1}$:
\begin{equation}
    {}^{\mathcal{Z}}\mb{\Phi}_{k+N_p}^{k+N_p+1} = \mb{T}_{k+N_p+1}^{-1}\mb{\Phi}_0^{k+N_p+1}\left(\mb{\Phi}_0^{k+N_p}\right)^{-1}\mb{T}_{k+N_p}.
\end{equation}

Note that
\begin{equation}
    \mb{\Phi}_0^{k+N_p+1}\left(\mb{\Phi}_0^{k+N_p}\right)^{-1} = \mb{\Phi}_0^{k+1}\mb{\Phi}_0^{N_p}\left(\mb{\Phi}_0^{N_p}\right)^{-1}\left(\mb{\Phi}_0^{k}\right)^{-1} = \mb{\Phi}_0^{k+1}\left(\mb{\Phi}_0^{k}\right)^{-1} = \mb{\Phi}_{k}^{k+1}.
\end{equation}

As a consequence
\begin{equation}
    {}^{\mathcal{Z}}\mb{\Phi}_{k+N_p}^{k+N_p+1} = \mb{\Gamma}^T\mb{T}_{k+1}^{-1}\mb{\Phi}_{k}^{k+1}\mb{T}_k\mb{\Gamma} = \mb{\Gamma}^{T} {}^{\mathcal{Z}}\mb{\Phi}_{k}^{k+1}\mb{\Gamma}.
\end{equation}

This means that ${}^{\mathcal{Z}}\mb{\Phi}_{k+N_p}^{k+N_p+1}$ and ${}^{\mathcal{Z}}\mb{\Phi}_{k}^{k+1}$ are related by the transformation defined by $\mb{\Gamma}$. In general,

\begin{equation}
    {}^{\mathcal{Z}}\mb{\Phi}_{k+mN_p}^{k+mN_p+1} = \left(\mb{\Gamma}^{T}\right)^m {}^{\mathcal{Z}}\mb{\Phi}_{k}^{k+1}\mb{\Gamma}^m.
\end{equation}

Moreover, since $\mb{R}_{\theta_\lambda}$ is the rotation matrix of $\theta_\lambda$ radians around the $z$-axis, each block diagonal element of $\mb{\Gamma}^m$ can be efficiently computed as the rotation of $m\theta_\lambda$ radians.

\section{Controller design}

\subsection{Problem statement}
\label{sec:problemStatement}
Assume we have a chaser spacecraft in the vicinity of the reference periodic orbit that admits an oscillatory mode. The goal is to design a controller that stabilizes the chaser to a desired reference trajectory on the invariant torus, which can be defined as a point $\mb{Z}_{ref} = [\mb{z}_{ref}; \dot{\mb{z}}_{ref}] = [\alpha_{ref}, \beta_{ref}, 0, 0, 0, 0]^T$ that lies within the center eigenspace of the reference orbit, as expressed in the toroidal coordinates. Because the toroidal frame co-rotates with the linear center mode, a fixed point in toroidal coordinates (constant $\mb{z}_{ref}$ with $\dot{\mb{z}}_{ref} = \mb{0}$) corresponds to the natural quasi-periodic relative motion on the invariant torus; the controller therefore regulates the chaser onto the natural quasi-periodic orbit rather than against it. The relative state of the chaser with respect to this reference trajectory in the toroidal frame is given by $\mb{\xi} = \mb{Z} - \mb{Z}_{ref}$.

The goal is to design a controller that stabilizes the chaser to the state $\mb{\xi} = \mb{0}$ while minimizing a certain cost function that depends on the control effort $\mb{u}$ and the state $\mb{\xi}$. Assume that the control inputs are impulses applied at the discrete time instants $t_k$. The dynamics of the system are given by the LTV system described in \cref{eq:dynamics}. Let the cost function be defined as
\begin{equation}
    J = \sum_{k=1}^{\infty}\mb{\xi}_k^T\mb{Q}\mb{\xi}_k + \mb{u}_k^T\mb{W}\mb{u}_k,
\end{equation}

where $\mb{Q}$ and $\mb{W}$ are positive definite matrices. It is proposed to design an LQR that minimizes the cost function $J$ subject to the system dynamics in \cref{eq:dynamics}.

\subsection{LQR problem resolution}

Finding the optimal control law $\mb{u}_k = -\mb{K}_k\mb{\xi}_k$ associated with the LQR problem defined in the previous section requires the stabilizing solution $\{\mb{P}_k\}$ of the backward difference Riccati equation (DRE)
\begin{equation}
    \mb{P}_k = \mb{A}_k^T\mb{P}_{k+1}\mb{A}_k - \mb{A}_k^T\mb{P}_{k+1}\mb{B}_k\left(\mb{W} + \mb{B}_k^T\mb{P}_{k+1}\mb{B}_k\right)^{-1}\mb{B}_k^T\mb{P}_{k+1}\mb{A}_k + \mb{Q}.
    \label{eq:DTARE}
\end{equation}

Note that \cref{eq:DTARE} is a recursion in $k$, not an \emph{algebraic} Riccati equation: the latter is the time-independent fixed point ($\mb{P}_k = \mb{P}_{k+1} = \mb{P}$) that exists only for time-invariant systems and yields a constant gain. For a general Linear Time-Varying system such as that described by \cref{eq:dynamics} no such steady state exists. While the infinite-horizon problem remains well-posed under uniform stabilizability and detectability, its stabilizing solution is an infinite, non-repeating sequence of matrices $\{\mb{P}_k\}$ that admits no finite parametrization and cannot, in general, be precomputed or stored for onboard implementation.

However, given the properties of $\mb{R}(t)$ described in the section on the time evolution of the transformation matrix, the matrices $\mb{A}_k$ and $\mb{B}_k$ satisfy the following relations:

\begin{equation}
    \mb{A}_{k+N_p} = \mb{\Gamma}^T\mb{A_k}\mb{\Gamma}, \quad \mb{B}_{k+N_p} = \mb{\Gamma}^T\mb{B}_k.
    \label{eq:RelationAforkNp}
\end{equation}

Let $\mb{Q}$ satisfy the invariance condition $\mb{\Gamma}^T\mb{Q}\mb{\Gamma} = \mb{Q}$, which holds, for example, when $\mb{Q}$ is a diagonal matrix whose entries for the $\alpha$ and $\beta$ coordinates are equal to each other and whose entries for the rates $\dot{\alpha}$ and $\dot{\beta}$ are equal to each other, e.g. $\mb{Q} = \text{diag}(q_1,q_1,q_2,q_{v1},q_{v1},q_{v2})$.

\textbf{Proposition.} Under the relations in \cref{eq:RelationAforkNp} and the invariance condition on $\mb{Q}$, the unique stabilizing solution $\{{\mb{P}_k}\}$ of the DRE~\cref{eq:DTARE} satisfies
\begin{equation}
    \mb{P}_{k+N_p} = \mb{\Gamma}^T\mb{P}_k \mb{\Gamma}.
    \label{eq:Pperiodicity}
\end{equation}

\textit{Proof.} Define the candidate sequence $\tilde{\mb{P}}_k := \mb{\Gamma}^T\mb{P}_k\mb{\Gamma}$ for all $k$. We show that $\{\tilde{\mb{P}}_k\}$ satisfies the DRE with system matrices $(\mb{A}_{k+N_p}, \mb{B}_{k+N_p})$. Substituting $\mb{A}_{k+N_p} = \mb{\Gamma}^T\mb{A}_k\mb{\Gamma}$, $\mb{B}_{k+N_p} = \mb{\Gamma}^T\mb{B}_k$, and $\tilde{\mb{P}}_{k+1} = \mb{\Gamma}^T\mb{P}_{k+1}\mb{\Gamma}$ into \cref{eq:DTARE}:
\begin{equation}
    \mb{A}_{k+N_p}^T\tilde{\mb{P}}_{k+1}\mb{A}_{k+N_p} = \mb{\Gamma}^T\mb{A}_k^T\underbrace{\mb{\Gamma}\mb{\Gamma}^T}_{=\mb{I}}\mb{P}_{k+1}\underbrace{\mb{\Gamma}\mb{\Gamma}^T}_{=\mb{I}}\mb{A}_k\mb{\Gamma} = \mb{\Gamma}^T\mb{A}_k^T\mb{P}_{k+1}\mb{A}_k\mb{\Gamma},
\end{equation}
where all $\mb{\Gamma}\mb{\Gamma}^T$ products reduce to $\mb{I}$ by orthogonality of $\mb{\Gamma}$. By the same mechanism, the cross term yields
\begin{equation}
\begin{split}
    &\mb{A}_{k+N_p}^T\tilde{\mb{P}}_{k+1}\mb{B}_{k+N_p}\left(\mb{W} + \mb{B}_{k+N_p}^T\tilde{\mb{P}}_{k+1}\mb{B}_{k+N_p}\right)^{-1}\mb{B}_{k+N_p}^T\tilde{\mb{P}}_{k+1}\mb{A}_{k+N_p}=\\
    &\quad=\mb{\Gamma}^T\mb{A}_k^T\mb{P}_{k+1}\mb{B}_k\left(\mb{W} + \mb{B}_k^T\mb{P}_{k+1}\mb{B}_k\right)^{-1}\mb{B}_k^T\mb{P}_{k+1}\mb{A}_k\mb{\Gamma}.
\end{split}
\end{equation}
Combining these and using $\mb{\Gamma}^T\mb{Q}\mb{\Gamma} = \mb{Q}$, the full recursion becomes
\begin{equation}
\begin{split}
    &\mb{\Gamma}^T\left[\mb{A}_k^T\mb{P}_{k+1}\mb{A}_k - \mb{A}_k^T\mb{P}_{k+1}\mb{B}_k\left(\mb{W} + \mb{B}_k^T\mb{P}_{k+1}\mb{B}_k\right)^{-1}\mb{B}_k^T\mb{P}_{k+1}\mb{A}_k + \mb{Q}\right]\mb{\Gamma} =\\
    &\quad= \mb{\Gamma}^T\mb{P}_k\mb{\Gamma} = \tilde{\mb{P}}_k.
\end{split}
    \label{eq:RiccatiKNp}
\end{equation}
Hence $\{\tilde{\mb{P}}_k\}$ satisfies the DRE for the shifted system $(\mb{A}_{k+N_p}, \mb{B}_{k+N_p})$. Under uniform stabilizability and detectability, the infinite-horizon Riccati recursion admits a unique bounded solution, which is positive semidefinite and stabilizing \cite{anderson1990optimal}. The candidate $\tilde{\mb{P}}_k = \mb{\Gamma}^T\mb{P}_k\mb{\Gamma}$ is bounded and positive semidefinite, since $\mb{P}_k$ is and $\mb{\Gamma}$ is orthogonal; hence it qualifies as that unique bounded solution. As $\{\mb{P}_{k+N_p}\}$ is precisely the bounded solution of the same shifted system (the shifted system inherits uniform stabilizability and detectability), it follows that $\mb{P}_{k+N_p} = \tilde{\mb{P}}_k = \mb{\Gamma}^T\mb{P}_k\mb{\Gamma}$.

As a direct consequence, the optimal control gain satisfies
\begin{equation}
    \mb{K}_{k+N_p} = \mb{K}_k\mb{\Gamma}.
    \label{eq:Kperiodicity}
\end{equation}

\Cref{eq:Pperiodicity,eq:Kperiodicity} imply that the control gains need only be computed offline for a single orbital period. The values at any subsequent time instant $k + mN_p$ are obtained by applying the orthogonal transformation $\mb{\Gamma}^m$, whose block-diagonal structure reduces this operation to a scalar rotation of $m\theta_\lambda$ radians. This enables the design of an infinite horizon LQR controller for the LTV system described by \cref{eq:dynamics}.

\section{Simulation results on CR3BP dynamics}
\label{sec:SimCR3BP}

In this section, the controller design is tested using the non-linear CR3BP dynamics. The objective is to perform a reconfiguration maneuver between two points that lie within the invariant curve at initial time.

\begin{figure}[htbp]
    \centering
    \includegraphics[width=0.8\textwidth]{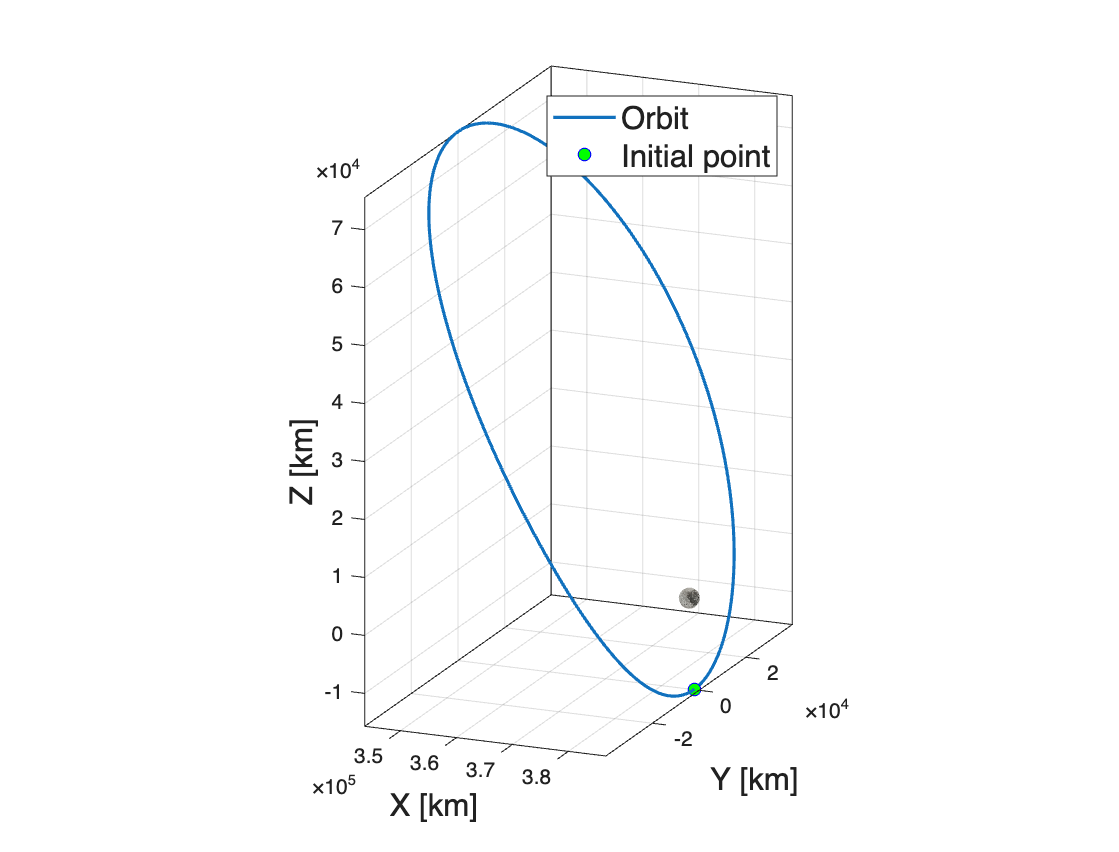}
    \caption{Reference orbit chosen for testing the designed LQR controller.}
    \label{fig:ref_trajectory}
\end{figure}

A reference orbit from the L1 Northern halo family with stability index close to unity is selected on the Earth-Moon system, with a period of $T = 9.33$ days. The STM and the transformation matrix $\mb{T}$ are computed for one full revolution of the orbit, with a time step of $\Delta t = 26.89$ minutes. The initial state for the reference trajectory on the rotating frame is given by $\mb{x}_0 = [0.989901409,   0.000784925,  0.040249211,   0.0019625251, -0.74435035,	-0.00791689]^T$ in non-dimensional units, starting near the perilune. The reference trajectory, together with the initial point are shown in \cref{fig:ref_trajectory}.

A value of $\epsilon$ is chosen so the chaser is initially at $1.12$ km from the reference orbit. The initial state is given by $\mb{Z}_{0} = [0,0.2566\times 10^{-4}, 0, 0, 0, 0]^T$. A reconfiguration is desired that moves the chaser to the target state $\mb{Z}_{ref} = [-0.222\times 10^{-4}, -0.1283\times 10^{-4}, 0, 0, 0, 0]^T$, which the controller regulates by driving $\mb{\xi} = \mb{Z} - \mb{Z}_{ref}$ to the origin. The controller is designed with time step equal to $\Delta t$. The cost matrices are defined as 

\begin{equation}
\mb{Q} = \text{diag}(10^{-2}, 10^{-2}, 10^{-2}, 10^{-3}, 10^{-3}, 10^{-3}), \quad \mb{W} = \text{diag}(1, 1, 1).
\end{equation}
 
\begin{figure}[htbp]
    \centering
    \includegraphics[width=0.9\textwidth]{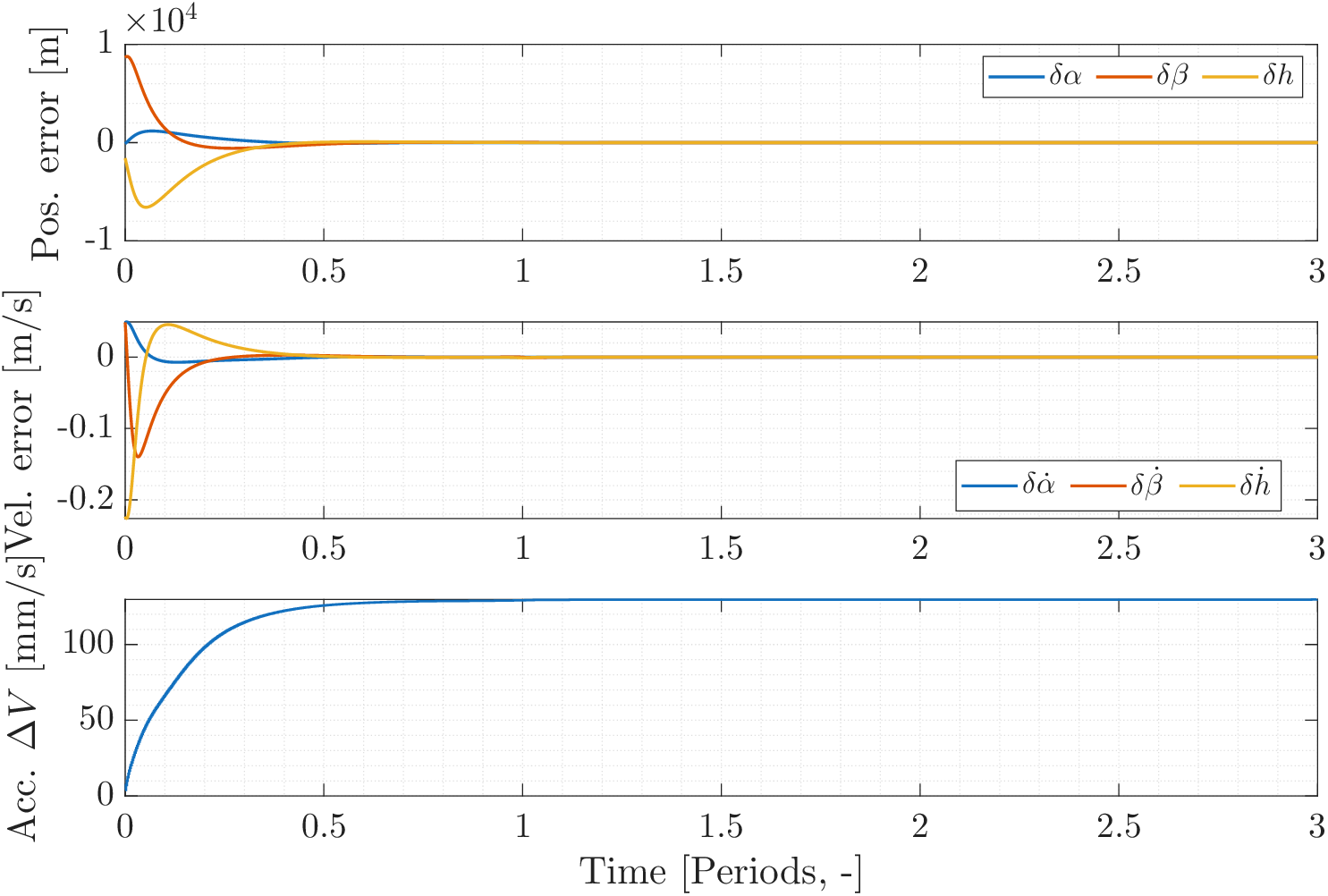}
    \caption{Evolution of the relative error with respect to the objective state on position (top) and velocity (middle). The bottom plot shows the control input applied at each time instant. Note that the control input is in mm/s.}
    \label{fig:LQR_evolution}
\end{figure}

\begin{figure}[htbp]
    \centering
    \includegraphics[width=0.8\textwidth]{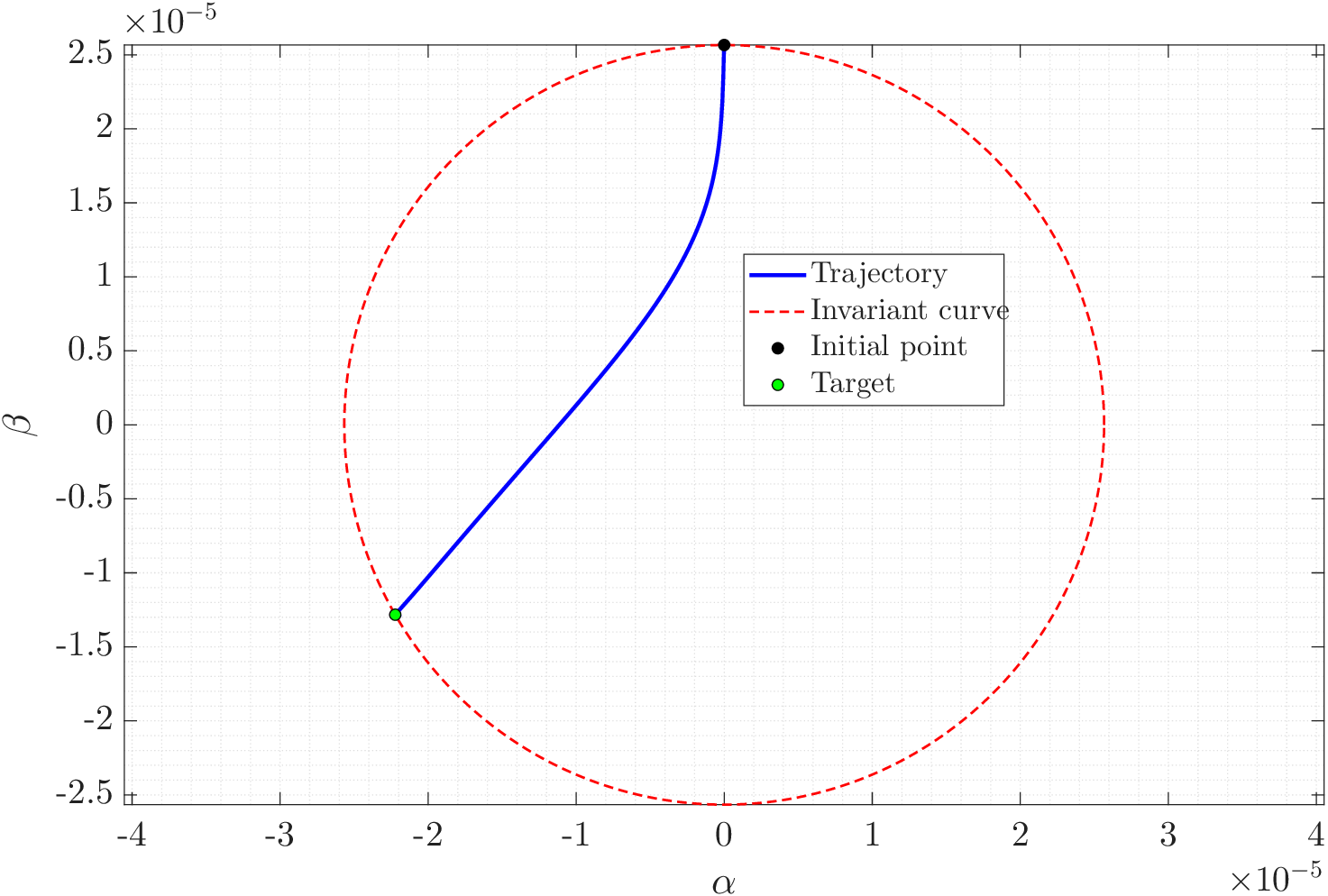}
    \caption{Invariant curve of the reference orbit in the $\alpha-\beta$ plane, together with the trajectory followed by the chaser on toroidal coordinates on the non-linear CR3BP model.}
    \label{fig:LQR_invariant}
\end{figure}

\Cref{fig:LQR_evolution} shows the time evolution of the relative position and velocity errors, as well as the accumulated control effort in mm/s. The designed controller is able to successfully drive the chaser to the desired state on the invariant curve. The total control effort for the maneuver is found to be below $130$ mm/s. Note that by tuning the values of $\mb{Q}$ and $\mb{W}$, it is possible to make the maneuver time longer or shorter, decreasing or increasing the control effort accordingly. \Cref{fig:LQR_invariant} shows the reconfiguration trajectory followed by the chaser in the $\alpha-\beta$ plane, together with the invariant curve of the reference orbit at the initial time.

\begin{figure}[htbp]
    \centering
    \includegraphics[width=1\textwidth]{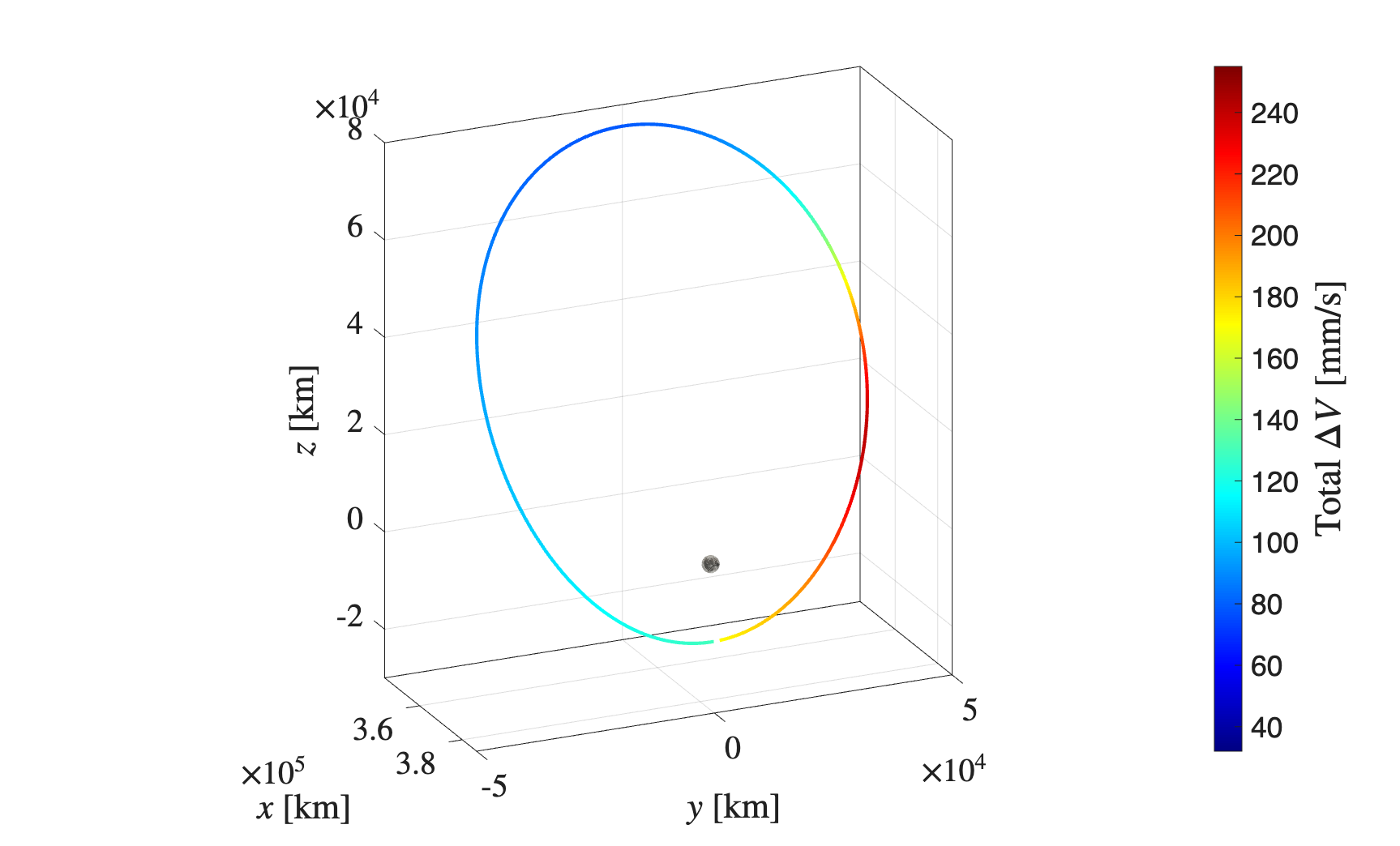}
    \caption{Total reconfiguration cost as a function of initial reference position within the reference orbit on the non-linear CR3BP model.}
    \label{fig:LQR_heatmap}
\end{figure}

Finally, \cref{fig:LQR_heatmap} shows the total control effort for the reconfiguration maneuver as a function of the initial reference position within the reference orbit. The tuning parameters and initial/final relative states within the toroidal frame are identical to the previous example. The cost is found to be higher, reaching up to approximately $260$ mm/s, when the initial position is close to but just before the perilune. This indicates that the transfer passes through the highly non-linear region near the Moon, which requires more control effort to stabilize the chaser. 

\section{Controller validation on ephemeris model}

In this section, the controller is tested using a fully non-linear ephemeris model of the Earth-Moon system, based on the NAIF SPICE toolkit \cite{SPICE1,SPICE2}, which provides the positions of the Earth and Moon in the J2000 inertial frame. The model also includes the gravitational perturbations of the Sun.

\subsection{Reference trajectory on the ephemeris model}

Since the periodic orbit is defined in the CR3BP, a multiple-shooting scheme is used to recover the trajectory that is closest to the original periodic orbit in the ephemeris model. The original trajectory is divided into $N$ arcs and used as the initial guess for the multiple-shooting scheme. The initial state, initial epoch and integration time of each arc are used as free variables. Continuity is imposed on the final state of each arc through the constraint vector. Additionally, the initial epoch is fixed, and the initial position is set to be on the $xz$-plane of the rotating frame. In order to improve numerical stability of the solution, the problem is non-dimensionalized using the reference length and time units of the CR3BP. Positiveness of the time of flight of each arc is ensured via a logarithmic transformation. 

The resulting problem is solved using Newton's method with minimum-norm update until the infinity-norm of the constraint vector is below a tolerance of $10^{-12}$. For the initial guess, 4 full periods of the reference CR3BP trajectory are used. The resulting points are  transformed between the rotating  and the inertial frame using the instantaneous relative positions of the Earth and Moon on the J2000 inertial frame at each epoch.

\begin{figure}[htbp]
    \centering
    \begin{minipage}[b]{0.495\textwidth}
        \centering
        \includegraphics[width=\textwidth]{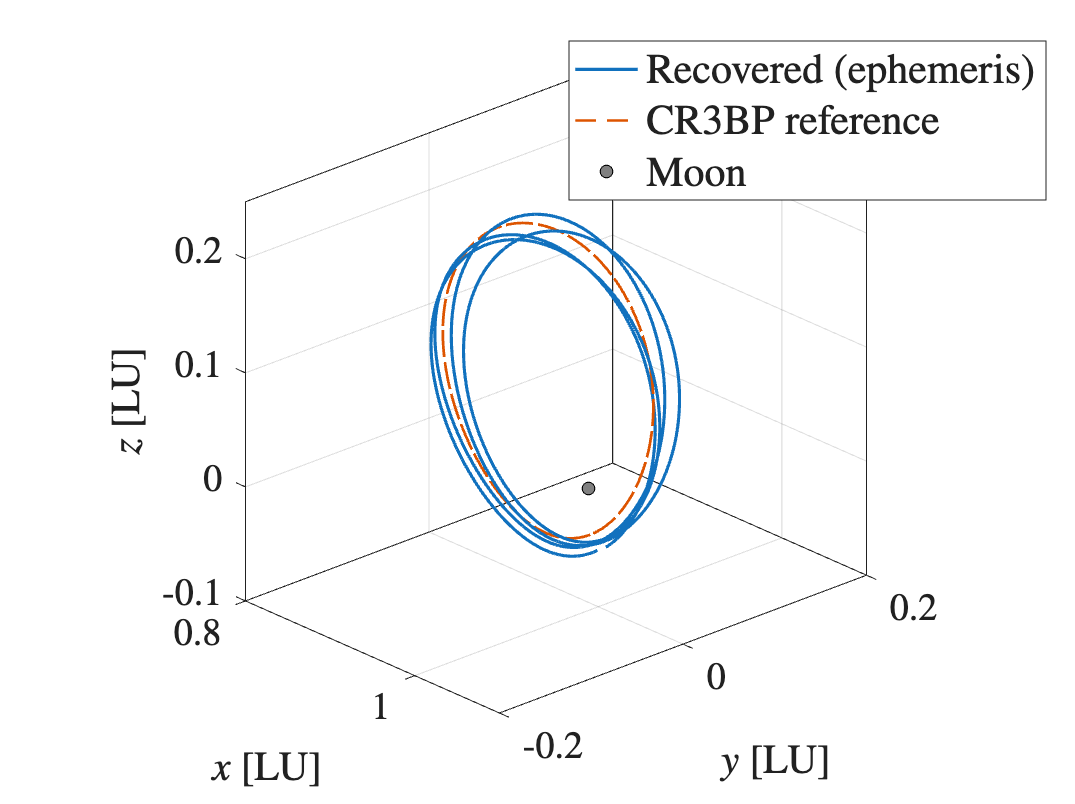}\\
        (a) Rotating frame.
    \end{minipage}
    \hfill
    \begin{minipage}[b]{0.495\textwidth}
        \centering
        \includegraphics[width=\textwidth]{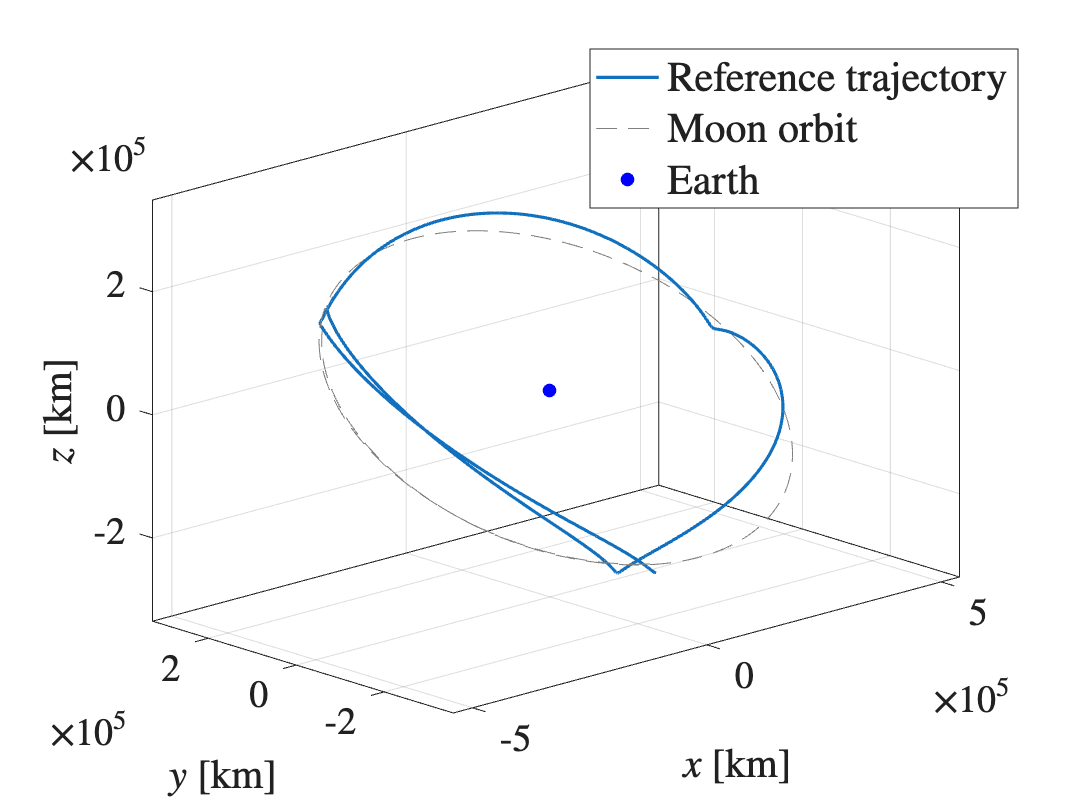}\\
        (b) Inertial frame.
    \end{minipage}
    \caption{Recovered trajectory for the ephemeris model at the reference epoch, shown in the (a) rotating and (b) inertial frames.}
    \label{fig:recoveredTrajectoryCombined}
\end{figure}

As a reference, the trajectory previously defined in the CR3BP is used. Since in the ephemeris model the system is no longer autonomous, it is necessary to define a reference epoch. In this case, the spacecraft is assumed to start at perilune at the epoch 2026-09-08T00:00:00.000 UTC. The resulting trajectory is shown in \cref{fig:recoveredTrajectoryCombined}, in both the rotating and inertial frames.

\subsection{Baseline controller}

As a reference for the evaluation of the performance of the proposed LQR controller, a combined transfer and station-keeping controller proposed in Reference~\citenum{elliott2021impulsive} is used. Given an initial state $\mb{Z}_0 = [\mb{z}_0; \dot{\mb{z}}_0]$ and a final desired state $\mb{Z}_f = [\mb{z}_f; \mb{0}]$ in toroidal coordinates and a transfer time $\tau$, the impulsive control action that performs the maneuver can be written as a function of ${}^{\mathcal{Z}}\mb{\Phi}(t_0+\tau; t_0)$ \cite{elliott2021impulsive}:

\begin{equation}
    \mb{u}_{trans,1} = \mb{R}(t_0)\left(\mb{\Phi}_{12}^{-1}\left(\mb{z}_f - \mb{\Phi}_{11}\mb{z}_0\right) - \dot{\mb{z}}_0\right),
\end{equation}

where $\mb{\Phi}_{11}$ and $\mb{\Phi}_{12}$ are the corresponding $3 \times 3$ submatrices of ${}^{\mathcal{Z}}\mb{\Phi}(t_0+\tau; t_0)$. Then, at $t = t_0 + \tau$, the second impulsive control action is applied to stop the relative motion, cancelling the toroidal relative velocity reached at the end of the transfer:

\begin{equation}
    \mb{u}_{trans,2} = -\mb{R}(t_0 + \tau)\dot{\mb{z}}(t_0+\tau).
\end{equation}

Once the transfer maneuver is completed, a station-keeping strategy is applied. This controller utilizes a first-order Taylor expansion of ${}^{\mathcal{Z}}\mb{\Phi}$, leading to a control action of the form:
\begin{equation}
    \mb{u}_{keep}  = -\mb{R}(t)\left(\dfrac{\mb{z}(t) - \mb{z}_{ref}}{\delta t} + \dot{\mb{z}}(t)\right),
\end{equation}
where $\delta t$ is the time between two consecutive impulsive control actions. The controller is implemented with an actuation interval of approximately 134 minutes ($0.01 T$). This combined targeting and station-keeping baseline is hereafter referred to as Targeting+SK.

\subsection{LQR controller performance on the ephemeris model}

Since the trajectory is not exactly periodic on the ephemeris model, it is necessary to have a continuous representation of the toroidal coordinates transformation matrix $\mb{T}$ and the gains $\mb{K}$ in the toroidal frame. This is achieved through a cubic spline interpolation of the eigenvectors of the monodromy matrix and the control gains during one full revolution of the CR3BP reference orbit, following a similar approach as in \cite{elliott2021impulsive}. For evaluating these polynomials, the time elapsed within the current revolution, normalized by the time between two consecutive crosses of the $xz$-plane with positive velocity along the rotating frame $y$-axis is used, defined as
\begin{equation}
    s_{curr} = \dfrac{t - t_p}{t_{p+1} - t_{p}},
\end{equation} 
where $p = 0\dots N_{rev}$ is the current revolution number, $t_p$ is the time of the last crossing of the $xz$-plane with positive velocity along the $y$-axis, and $t_{p+1}$ is the time of the next crossing. Additionally, the control gain for any given time instant $t$ is computed as
\begin{equation}
    \mb{K}(t) = \mb{K}(s_{curr})\mb{\Gamma}^{p},
\end{equation}
taking into account the rotation through matrix $\mb{\Gamma}$ for each revolution completed.

The weight matrices of the LQR controller are set to
\begin{equation}
\mb{Q} = \text{diag}(10^2, 10^2, 10^2, 10^{-1}, 10^{-1}, 10^{-1}), \quad \mb{W} = \text{diag}(1, 1, 1)\times 10^{-2}.
\label{eq:refQmatrix}
\end{equation}

The controllers are tested around the reference trajectory recovered on the ephemeris model by performing a reconfiguration maneuver between the very same two points on the invariant curve as in the CR3BP simulation results presented previously.

\begin{figure}[htbp]
    \centering
    \includegraphics[width=0.9\textwidth]{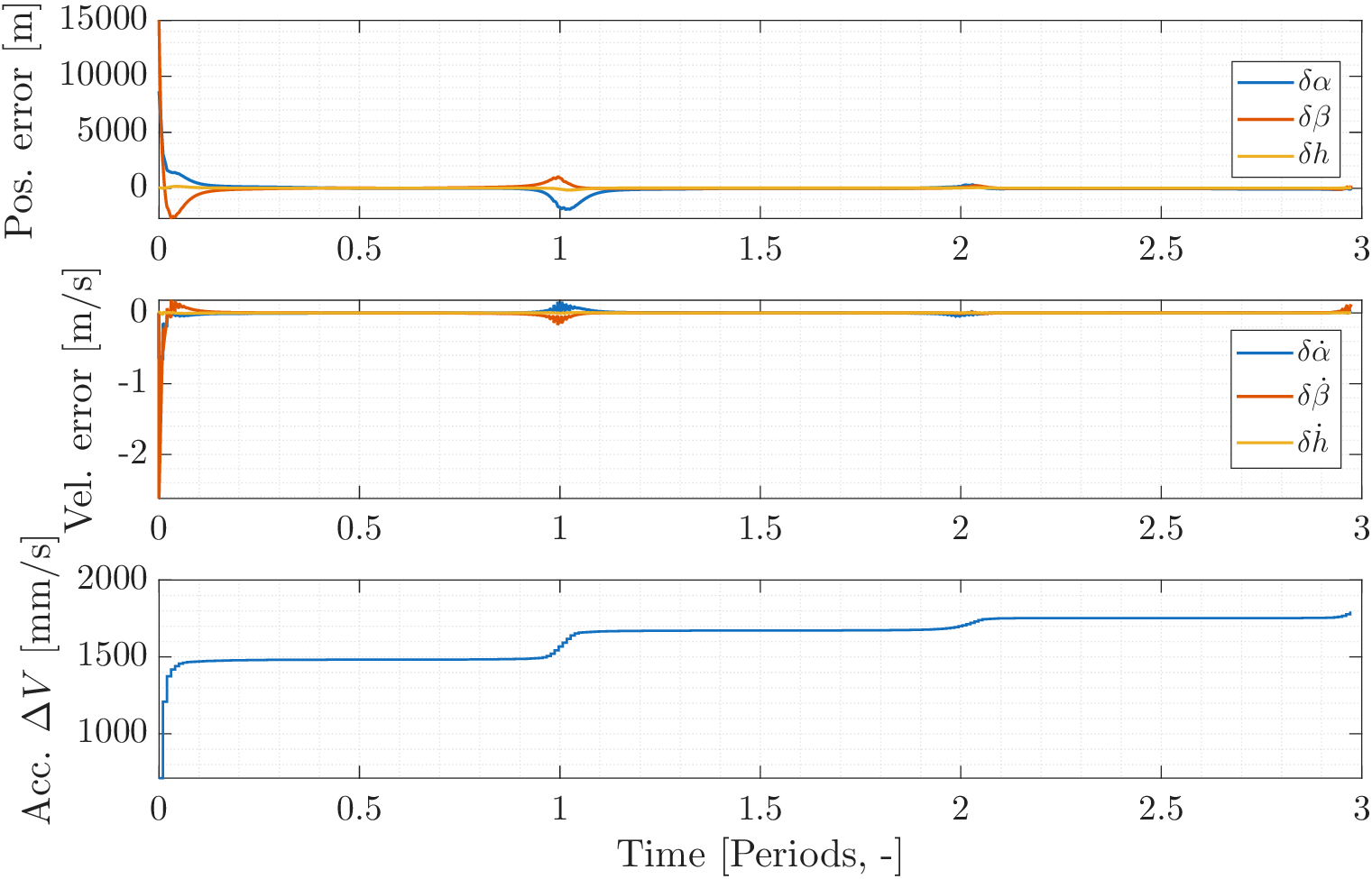}
    \caption{Evolution of the relative error with respect to the objective state on position (top) and velocity (middle) for the ephemeris model simulation. The bottom plot shows the control input applied at each time instant. Note that the control input is in mm/s.}
    \label{fig:LQR_eph_evolution}
\end{figure}

\Cref{fig:LQR_eph_evolution} shows the time evolution of the relative position and velocity errors, together with the cumulative control effort applied. At perilune, the high non-linearity of the dynamics produces some oscillations on the first passage that are reduced on subsequent revolutions. Note that the weight matrix $\mb{Q}$ gives more weight to the position error than in the CR3BP simulation results presented previously. This more aggressive tuning is motivated by the larger discrepancies between the ephemeris model and the linearized CR3BP dynamics on which the LQR is designed: it is found to greatly reduce the magnitude of the perilune oscillations, keeping the errors below 1 km at the first perilune passage, at the cost of a higher control effort.

The higher position weight also produces faster convergence of the controller (with errors below 500 m after approximately 20 hours). The total $\Delta V$ required is on the order of 1.79 m/s. Calibrating the LQR controller of the CR3BP simulation results using a more aggressive tuning to obtain equivalent convergence times requires a total $\Delta V$ on the order of 0.6 m/s. This difference in control cost is again attributed to these discrepancies.

\subsection{Controller comparison}

\begin{figure}[htbp]
    \centering
    \includegraphics[width=0.9\textwidth]{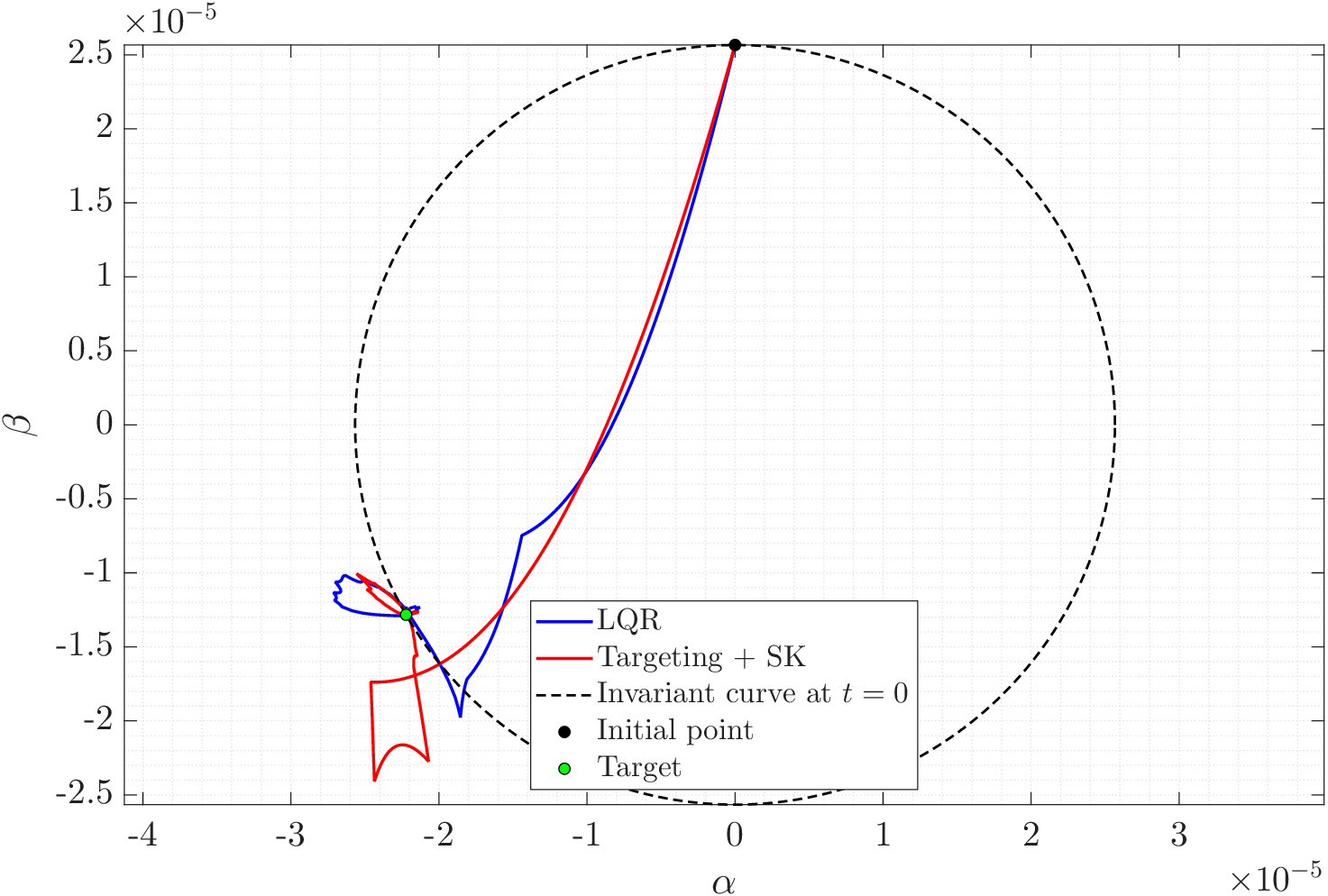}
    \caption{Trajectory followed on the $\alpha-\beta$ plane for the ephemeris model simulation, together with the invariant curve at initial time.}
    \label{fig:LQR_eph_invariant}
\end{figure}

\Cref{fig:LQR_eph_invariant} shows the trajectory followed by the chaser in the $\alpha-\beta$ plane for the LQR and the baseline controller (Targeting+SK), together with the invariant curve at initial time. Both controllers are able to reach the target state on the invariant curve. Some oscillations are found near the target, which can be linked with the evolution found in \cref{fig:LQR_eph_evolution}. The baseline controller also suffers from this behavior after the transfer, but is likewise able to reduce them and stabilize the position near the reference.

\begin{figure}[htbp]
    \centering
    \includegraphics[width=0.9\textwidth]{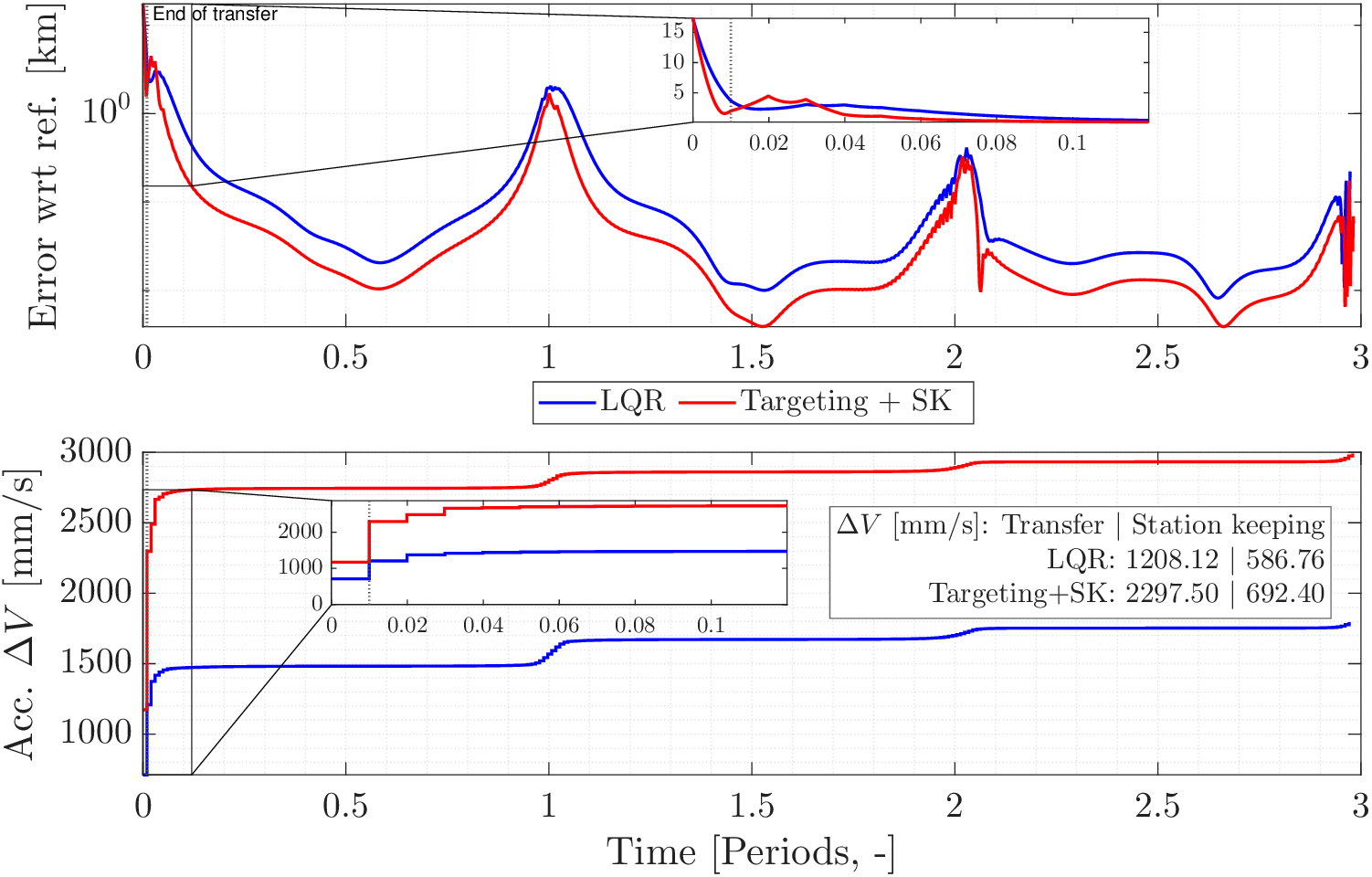}
    \caption{Position error comparison between the LQR and the baseline controller (top) and accumulated control effort comparison (bottom) for the ephemeris model simulation.}
    \label{fig:LQR_eph_evolution_comparison}
\end{figure}

\Cref{fig:LQR_eph_evolution_comparison} compares the LQR and the baseline controller. The top plot shows the position error on a logarithmic scale, and the bottom plot the accumulated control effort. Both controllers keep the Root-Mean-Square (RMS) error below 400 m after the first revolution, with the baseline reaching a somewhat lower error. However, the LQR achieves this with a total $\Delta V$ of 1.79 m/s against 2.99 m/s for the baseline, a 40\% reduction in control effort. \Cref{tab:error_after_first_rev} summarizes the performance of both controllers. The larger maximum errors reported in \cref{tab:error_after_first_rev} (around 2 km) correspond to transient spikes at the recurring perilune passages, where the dynamics are most non-linear; between passages the error remains below 500 m. 

\begin{table}[htbp]
\centering
\caption{Total control effort and error with respect to reference after the first revolution ($t > 1$ period).}
\begin{tabular}{|l|c|c|c|}
\hline
\textbf{Controller} & \textbf{RMS [km]} & \textbf{Max [km]} & \textbf{Total $\Delta V$ [m/s]} \\
\hline
LQR                 & 0.3938            & 2.0595            & 1.79                              \\
Targeting+SK        & 0.1738            & 1.6870            & 2.99                             \\
\hline
\end{tabular}
\label{tab:error_after_first_rev}
\end{table}

\subsection{Sensitivity analysis on weight matrices}

Note that the LQR performance is sensible to the values set in the weight matrices $\mb{Q}$ and $\mb{W}$. A more aggressive tuning of the weights can reduce the maximum error at the cost of a higher control effort.

\begin{figure}[htbp]
    \centering
    \includegraphics[width=0.92\textwidth]{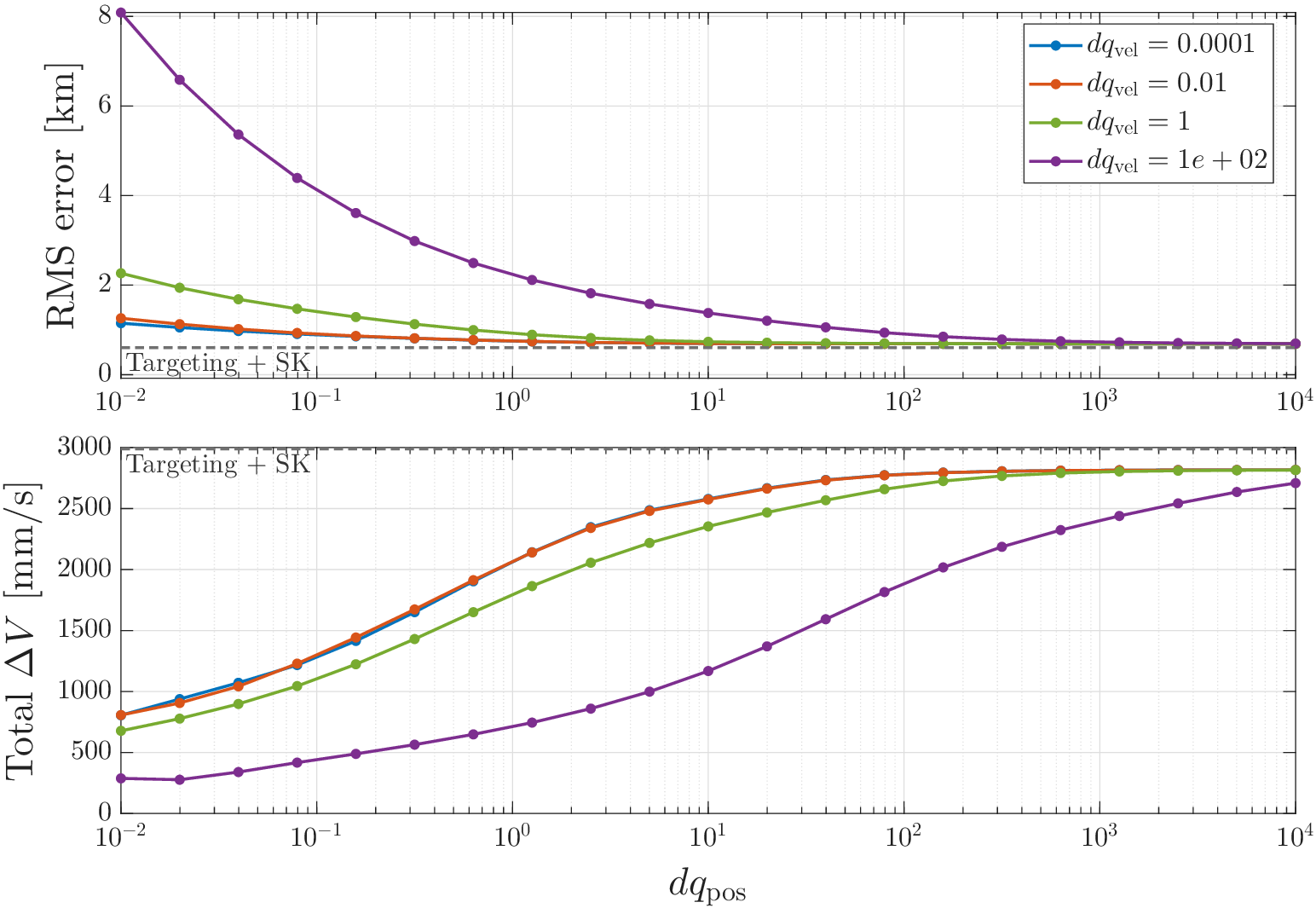}
    \caption{Performance evolution of the LQR controller on the ephemeris model as a function of variations on the values of the weight $\mb{Q}$.}
    \label{fig:LQR_eph_sensitivity}
\end{figure}

In order to analyze the sensitivity to the weight matrix value, let $\mb{Q}_{par}$ be a diagonal weight matrix
\begin{equation}
    \mb{Q}_{par} = \begin{bmatrix}
        dq_p \mb{I}_3 & 0 \\
        0 & dq_v \mb{I}_3
    \end{bmatrix} \mb{Q},
\end{equation}
where $dq_p$ and $dq_v$ are the position and velocity weights with respect to the original matrix $\mb{Q}$ defined in \cref{eq:refQmatrix}. Matrix $\mb{W}$ is left as in the previous section. The total control effort and the RMS position error are computed for different values of $dq_p$ and $dq_v$. \Cref{fig:LQR_eph_sensitivity} shows the results of this analysis. As expected, increasing the value of $dq_p$ reduces the RMS error, while increasing the total control cost. On the other hand, increasing the value of $dq_v$ reduces the error performance and the control effort. This is because the velocity weight indirectly limits the control action, as the controller is penalized for applying a control that produces a high relative velocity to the reference point. Note that tuning appropiately the values of $dq_p$ and $dq_v$ it is possible to obtain performances very close to the baseline controller, while still reducing the total control effort.

\section{Conclusions}

This paper presents the design of an LQR controller for formation flying near periodic orbits within the CR3BP, formulated in non-singular toroidal coordinates. The primary contribution lies in showing that the rotational quasi-periodicity of the toroidal transformation matrix $\mb{T}$, captured by the orthogonal matrix $\mb{\Gamma}$, reduces the infinite-horizon difference Riccati equation for the resulting Linear Time-Varying system to a single-period computation. This was established through a uniqueness argument on the stabilizing solution of the resulting quasi-periodic Riccati recursion, allowing both the Riccati matrices and the control gains to be computed offline for one orbital period and extended to arbitrary time horizons via a simple rotation.

A first validation against the full non-linear CR3BP dynamics on an L1 Northern halo orbit in the Earth-Moon system indicates that the controller can drive a chaser spacecraft to a target state on the invariant curve with a total $\Delta V$ budget below $260$ mm/s even when starting the maneuver just before arriving at the perilune.

The controller has been further validated against a full ephemeris model of the Earth-Moon system, built on the NAIF SPICE toolkit and including the solar gravitational perturbation. A reference quasi-periodic trajectory was recovered in this model through a multiple-shooting scheme, and the gains designed on the CR3BP were extended to the non-periodic ephemeris dynamics by interpolating the toroidal transformation over one revolution and applying the $\mb{\Gamma}$ rotation at each subsequent revolution. The controller successfully drives the chaser to the target state on the invariant curve, keeping the position error below $400$ m (RMS) after the first revolution; the larger excursions are transient spikes confined to the highly non-linear perilune passages, which can be attenuated through a more aggressive tuning of the weight matrices.

Finally, the proposed LQR has been compared against an impulsive targeting plus station-keeping controller taken as a baseline. Both controllers achieve a comparable tracking accuracy, but the LQR attains it with a total $\Delta V$ of $1.79$ m/s against $2.99$ m/s for the baseline, a reduction of approximately $40\%$ in control effort. This indicates that the proposed formulation is competitive with existing impulsive strategies, while retaining the optimality and the compact onboard representation afforded by the quasi-periodic symmetry.

For future work, it is proposed to apply the developed controller for the design of Model Predictive Control strategies with stability guarantees by introducing terminal ingredients \cite{mayne2000survey}.

\section{Acknowledgements}
The authors acknowledge partial support by grant PID2023-147623OB-I00 funded by\\ MICIU/AEI/10.13039/501100011033 and by ``ERDF A way of making Europe." Travel to this conference was supported by AFOSR/EOARD grant FA8655-25-1-7012.
\bibliographystyle{AAS_publication}   
\bibliography{main}   

@article{elliott2022describing,
  title={Describing relative motion near periodic orbits via local toroidal coordinates},
  author={Elliott, Ian and Bosanac, Natasha},
  journal={Celestial Mechanics and Dynamical Astronomy},
  volume={134},
  number={2},
  pages={19},
  year={2022},
  publisher={Springer}
}

@article{capannolo2023model,
  title={Model predictive control for formation reconfiguration exploiting quasi-periodic tori in the cislunar environment},
  author={Capannolo, Andrea and Zanotti, Giovanni and Lavagna, Mich{\`e}le and Cataldo, Giuseppe},
  journal={Nonlinear Dynamics},
  volume={111},
  number={8},
  pages={6941--6959},
  year={2023},
  publisher={Springer}
}

@inproceedings{elliott2021impulsive,
  title={Impulsive control of formations near invariant tori via local toroidal coordinates},
  author={Elliott, Ian and Bosanac, Natasha},
  booktitle={Proceedings of the AAS/AIAA Astrodynamics Specialist Virtual Conference, Online},
  pages={9--12},
  year={2021}
}

@inproceedings{lopez2022control,
  title={Control design for rendezvous operation near Halo Orbits using Lyapunov-Floquet theory and AUTO},
  author={L{\'o}pez-Cepero, Joaquin G and Galan-Vioque, Jorge and Vazquez, Rafael},
  booktitle={CEAS EuroGNC 2024},
  year={2022}
}

@inproceedings{howell2018nrho,
  title={Earth-Moon Near Rectilinear Halo and Butterfly Orbits for Lunar Surface Exploration},
  author={Whitley, Ryan and Davis, D. and Burke, L. and Howell, Kathleen C.},
  booktitle={AAS/AIAA Astrodynamics Specialist Conference},
  year={2018}
}

@book{szebehely1967theory,
  title={Theory of Orbits: The Restricted Problem of Three Bodies},
  author={Szebehely, Victor},
  year={1967},
  publisher={Academic Press}
}

@article{sanchez2020chance,
  title={Chance-constrained model predictive control for near rectilinear halo orbit spacecraft rendezvous},
  author={Sanchez, Julio C and Gavilan, Francisco and Vazquez, Rafael},
  journal={Aerospace Science and Technology},
  volume={100},
  pages={105827},
  year={2020},
  publisher={Elsevier}
}

@article{mayne2000survey,
  title={Survey constrained model predictive control: Stability and optimality},
  author={Mayne, David Q and Rawlings, James B and Rao, Christopher V and Scokaert, Pierre OM},
  journal={Automatica (Journal of IFAC)},
  volume={36},
  number={6},
  pages={789--814},
  year={2000},
  publisher={Pergamon Press, Inc. Elmsford, NY, USA}
}

@article{peng2011optimal,
  title={Optimal periodic controller for formation flying on libration point orbits},
  author={Peng, Haijun and Zhao, Jun and Wu, Zhigang and Zhong, Wanxie},
  journal={Acta Astronautica},
  volume={69},
  number={7-8},
  pages={537--550},
  year={2011},
  publisher={Elsevier}
}

@inproceedings{whitley2016options,
  title={Options for staging orbits in cislunar space},
  author={Whitley, Ryan and Martinez, Roland},
  booktitle={2016 IEEE Aerospace Conference},
  pages={1--9},
  year={2016},
  organization={IEEE}
}

@article{SPICE1,
  title={Ancillary data services of NASA's navigation and ancillary information facility},
  author={Acton Jr, Charles H},
  journal={Planetary and Space Science},
  volume={44},
  number={1},
  pages={65--70},
  year={1996},
  publisher={Elsevier}
}

@article{SPICE2,
  title={A look towards the future in the handling of space science mission geometry},
  author={Acton, Charles and Bachman, Nathaniel and Semenov, Boris and Wright, Edward},
  journal={Planetary and Space Science},
  volume={150},
  pages={9--12},
  year={2018},
  publisher={Elsevier}
}

@book{anderson1990optimal,
  title={Optimal Control: Linear Quadratic Methods},
  author={Anderson, Brian D. O. and Moore, John B.},
  publisher={Prentice-Hall},
  address={Englewood Cliffs, NJ},
  year={1990}
}

\end{document}